\documentclass[a4paper]{jpconf}
\usepackage{graphicx}
\usepackage{amsmath}
\usepackage{amssymb}

\def\Msusy{m_{\rm stop}}
\newcommand{\vev}[1]{ \left\langle {#1} \right\rangle }

\begin{document}
\title{Natural supersymmetric Twin Higgs}

\author{Marcin Badziak$^1$ and Keisuke Harigaya$^2$}

\address{$^1$ Institute of Theoretical Physics, Faculty of Physics, University of Warsaw, ul.~Pasteura 5, PL--02--093 Warsaw, Poland}

\address{$^2$ Institute for Advanced Study, Natural Sciences, 1 Einstein Drive, Princeton, New Jersey 08540, United States}

\ead{mbadziak@fuw.edu.pl}

\begin{abstract}
A new class of supersymmetric Twin Higgs (TH) models where new gauge symmetry is responsible for the TH mechanism is reviewed. In this class of models the Higgs mass is naturally in agreement with the LHC measurement while the electroweak symmetry breaking is realised without excessive tuning despite of strong lower bounds on masses of supersymmetric particles set by the LHC. Assuming particular non-abelian structure of the new gauge symmetry the model remains perturbative up to the energy scale of gravity, in contrast to all previously proposed UV completions of the TH model.  
\end{abstract}

\section{Introduction}

The lack of finding any new particles beyond the Standard Model (SM) at the LHC has substantially increased the fine-tuning of the electroweak symmetry breaking (EWSB) in models addressing the hierarchy problem of the SM such as supersymmetry (SUSY) or Composite Higgs.  This is indication of the little hierarchy problem.

A possible solution to the little hierarchy problem is provided by models of Neutral Naturalness in which top partners responsible for the cancellation of quadratic divergences to the Higgs mass parameter do not carry the SM color charge. The experimental constraints on such top partners are strongly relaxed making the fine-tuning of the EW scale much smaller. 

Arguably the most widely studied model of Neutral Naturalness is the Twin Higgs (TH) model~\cite{Chacko:2005pe,Chacko:2005vw,Chacko:2005un,Falkowski:2006qq,Chang:2006ra}. In this scenario, the Higgs is a pseudo-Nambu-Goldstone boson of a global $SU(4)$ symmetry emerging from $\mathbb{Z}_2$ symmetry exchanging the SM with its mirror (or twin) copy. We denote mirror objects with supersctripts $'$. The Higgs potential in this class of models can be quite generally written in the following way:
\begin{align}
V = \lambda (|H'|^2 + |H|^2)^2 -m^2 (|H'|^2 + |H|^2) + \Delta \lambda(|H'|^4 + |H|^4) + \Delta m^2 |H^2| \,,
\end{align}
where $H$ is the SM Higgs doublet while $H'$ its mirror counterpart.  The first two terms are both $\mathbb{Z}_2$ and $SU(4)$ symmetric, $\Delta \lambda$ preserves $\mathbb{Z}_2$ but breaks $SU(4)$ which is necessary to make the Higgs massive. Since the LHC Higgs measurements show that the discovered Higgs has properties similar to the SM Higgs $\mathbb{Z}_2$ must be broken which is parameterised by $\Delta m^2$ in the above potential. This $\mathbb{Z}_2$ breaking introduces some tuning in the potential:
\begin{equation}
\Delta_{v/f} = \frac{1}{2} \left( \frac{f^2}{v^2} -2\right) \,,
\end{equation}
where $\vev{H} \equiv v$, $\vev{H'}\equiv v'$, and $f \equiv \sqrt{v^2 + v^{'2}}$ is the decay constant of the spontaneous $SU(4)$ breaking.
However, currently this results only in a minor tuning of $\mathcal{O}(20\%)$ because the LHC Higgs data set a constraint $f\gtrsim3v$ \cite{Higgscomb}. 

It should be emphasised that the TH model solves only the little hierarchy problem. A fully satisfactory TH model should be UV completed in a way that solve also the big hierarchy problem of the SM. Existing UV completions  of the TH model involve either Composite Higgs~\cite{Batra:2008jy,Geller:2014kta,Barbieri:2015lqa,Low:2015nqa} or supersymmetry~\cite{Falkowski:2006qq,Chang:2006ra,Craig:2013fga,Katz:2016wtw,Badziak:2017syq,Badziak:2017kjk,Badziak:2017wxn}. The UV completed TH models have additional source of tuning on top of that from $Z_2$ breaking mentioned above.  This is because in UV completed models ordinary top partners that carry QCD charge must be introduced to solve the big hierarchy problem. Due to the TH mechanism the tuning resulting from introducing such states  is suppressed by a factor $\lambda_{\rm SM}/(2\lambda)$ (where $\lambda_{\rm SM}\approx0.13$ is the SM Higgs quartic coupling) as compared to corresponding models without the TH mechanism. In any UV completion the $SU(4)$ invariant coupling $\lambda$ cannot be arbitrarily large so color charged top partners may introduce non-negligible tuning if they are too heavy. 

In the context of SUSY UV completions of the TH model the upper bound on $\lambda$, hence the lower bound on the fine-tuning, originates from perturbativity constraints.  Another constraint on SUSY TH models that may impact the fine-tuning comes from the Higgs mass measurement.  In initially proposed SUSY UV completions of the TH model  in which $\lambda$ is generated from an $F$-term potential of some new singlet~\cite{Falkowski:2006qq,Chang:2006ra,Craig:2013fga} these constraints make the TH mechanism totally inefficient and the fine-tuning is not improved with respect to SUSY models without TH mechanism~\cite{Craig:2013fga,Badziak:2017syq}.

In these proceedings we review a new class of SUSY TH models in which the $SU(4)$ invariant quartic term is generated by a $D$-term potential of a new gauge symmetry~\cite{Badziak:2017syq,Badziak:2017kjk,Badziak:2017wxn}. In this class of models the Higgs mass constraint is easily satisfied even for light stops while EWSB does not require excessive fine-tuning.

\section{SUSY $U(1)_X$ $D$-term Twin Higgs}

We start with a model in which a large $SU(4)$ invariant quartic term originates from a
non-decoupling $D$-term of a new $U(1)_X$ gauge symmetry~\cite{Badziak:2017syq}. 
Such a non-decoupling $D$-term may be present if the mass of a scalar field responsible for
the breaking of the $U(1)_X$ gauge symmetry is dominated by a SUSY breaking soft mass. 
We introduce chiral multiplets
$\Xi$, $S$ and $\bar{S}$ whose $U(1)$ charges are $0$, $+q$ and $-q$, respectively, and the superpotential,
\begin{align}
\label{eq:W_S}
W = \kappa \Xi (S\bar{S}- M^2),
\end{align}
where $\kappa,~M$ are constants,
and soft masses,
\begin{align}
\label{eq:soft_S}
V_{\rm soft} = m_S^2 (|S|^2 + |\bar{S}|^2  ).
\end{align}
Here we assume that the soft masses of $S$ and $\bar{S}$ are the same. Otherwise, the asymmetric VEVs of $S$ and $\bar{S}$ give a large soft mass to
the Higgs doublet through the D-term potential of $U(1)_X$.

Assuming that all Higgs bosons are charged under the new $U(1)_X$ gauge symmetry and  integrating out $S$ fields 
we obtain the non-decoupling $D$-term potential:
\begin{align}
\label{eq:VU1X}
 V_{U(1)_X}=\frac{g_X^2}{8} \left( |H_u|^2-|H_d|^2 + |H'_u|^2-|H'_d|^2 \right)^2 \left(1-\epsilon^2\right) \,, ~~
\epsilon^2 \equiv \frac{m_X^2}{2m_S^2 + m_X^2},
\end{align}
where $m_X^2= 4 g_X^2 q^2 v_S^2$ is the $U(1)_X$ gauge boson mass with $g_X$ the $U(1)_X$  gauge coupling and $v_S$ the VEV of $S$ and $\bar{S}$.
This term gives the following $SU(4)$ invariant coupling:
\begin{equation}
\label{eq:lambdaD}
 \lambda=g_X^2\frac{\cos^2\left( 2\beta \right)}{8}\left(1-\epsilon^2\right) \,.
\end{equation}
Note that $\lambda$ is maximized in the limit of large $\tan\beta$ which makes it easier to satisfy
the Higgs mass constraint. Large $\lambda$ prefers also $g_X$ as large as possible and $\epsilon\ll1$. 

The magnitude of $g_X$ is constrained from above by perturbativity.  The beta function of the 
$U(1)_X$ gauge coupling constant depends on the charge assignment of particles in the visible and mirror sectors. It turns out that the beta function is minimized when the $U(1)_X$ charges of the MSSM
particles and the mirror particles are the following linear combination of $U(1)_Y$ and $U(1)_{B-L}$ charges:
\begin{equation}
q_X = q_Y -\frac12 q_{\rm B-L} \,.
\end{equation}
Then the beta function of the $U(1)_X$ gauge coupling constant is given by
\begin{equation}
\frac{\rm d}{{\rm dln} \mu}\frac{8\pi^2}{g_X^2} = b_X,~~ b_X =
\begin{cases}
- 14 & \text{mirror} \\
-10 & \text{fraternal}
\end{cases}
\end{equation}
where in fraternal Twin Higgs models~\cite{Craig:2015pha} the first and the second generation fermions do not have their twins so the RG running is slightly slower.

The scale of the Landau pole $M_{c}$ is given by
\begin{equation}
M_c = m_X \times {\rm exp} [  - \frac{8\pi^2}{b_Xg_X(m_X)^2}].
\end{equation}
We require that $M_c$ is larger than the mediation scale of the SUSY breaking,
which is typically at the smallest around $100 m_{\rm stop}$. 
The electroweak precision as well as the production of di-muon at the LEP put a lower bound $m_X/ g_X \gtrsim 4$ TeV.
This requires $M_c \gtrsim 10m_X$ which sets an upper bound on $g_X(m_X)$ of
about 1.6 (1.9) for the mirror (fraternal) Twin Higgs model.

\begin{figure}[t]
\centering
\includegraphics[width=.49\textwidth]{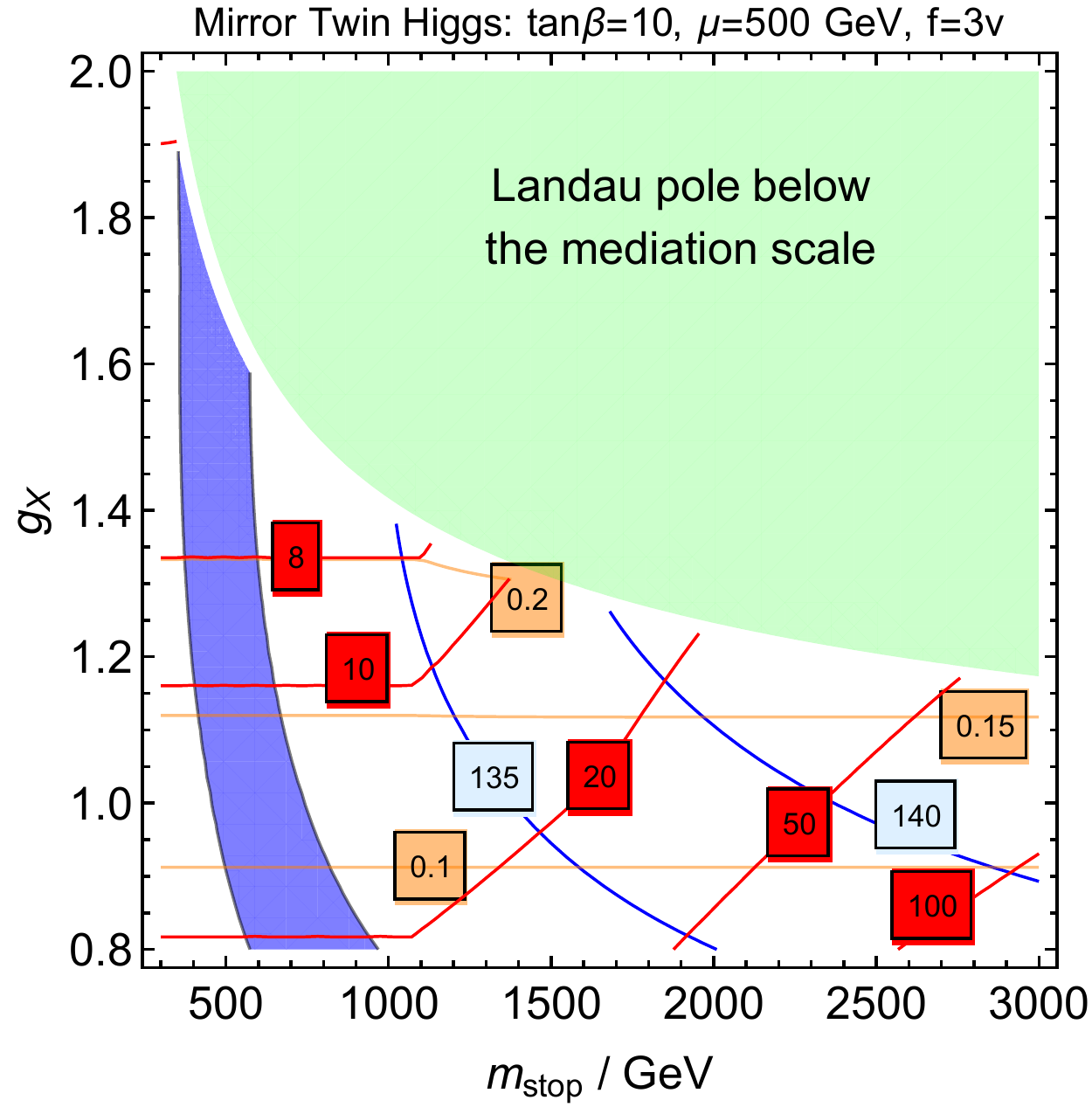}
\includegraphics[width=.49\textwidth]{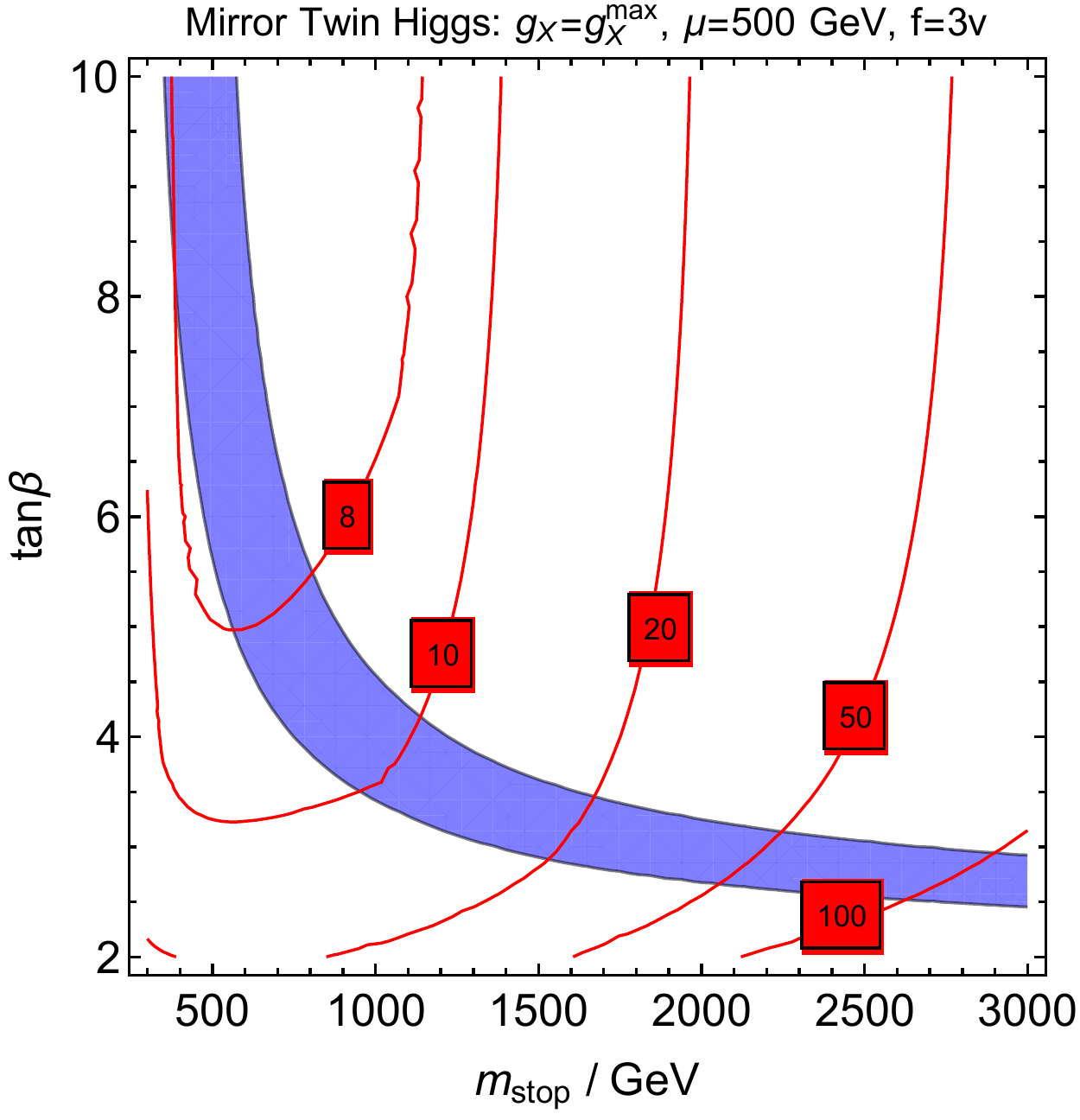}
\caption{Fine-tuning (red contours) in the $U(1)_X$ $D$-term  Twin Higgs model for $f=3v$, $\mu=500$ GeV, $m_A=1$ TeV and $M_3=2$ TeV  assuming the messenger
scale $\Lambda=100 \Msusy$. In the left panel, where $\tan\beta=10$, the orange contours depict the value of the $SU(4)$ preserving quartic coupling
and in the green regions the Landau pole of the $U(1)_X$ gauge coupling constant is below $\Lambda$. In the right panel, at each point of the plane
$\Msusy$-$\tan\beta$, $g_X$ is fixed to the maximal value that allows the messenger scale to be below the Landau pole. In the blue region the Higgs
mass is in agreement with the measured value and several blue contours of the Higgs mass are also shown. 
}
\label{fig:Dterm}
\end{figure}

In order to quantify the fine-tuning of EWSB we use the following measure:
\begin{equation}
\label{eq:Delta_v}
\Delta_v \equiv   \Delta_f \times \Delta_{v/f},
\end{equation}
where the tuning in percent is $100\%/\Delta_v$ and
\begin{align}
\label{eq:Delta_vf}
\Delta_{v/f} = \frac{1}{2} \left( \frac{f^2}{v^2} -2\right),
\Delta_f =  {\rm max}_i \left( |\frac{\partial{\rm ln} f^2}{\partial{\rm ln} x_i(\Lambda)}|, 1 \right) .
\end{align}
$\Delta_{v/f}$ measures the tuning to obtain $v < f$
via explicit soft $\mathbb{Z}_2$ symmetry breaking which is required by the Higgs coupling measurements~\cite{Higgscomb}. $\Delta_f$ measures the tuning to obtain the scale $f$
from the soft SUSY breaking. $x_i(\Lambda)$ are the parameters of the theory evaluated at the mediation scale of the SUSY breaking $\Lambda$. More details about our procedure to calculate $\Delta_v$ can be found in~\cite{Badziak:2017syq,Badziak:2017kjk,Badziak:2017wxn}. Apart from the usual tuning from stops, the tuning may also
arise from a threshold correction to the soft Higgs mass which is proportional to
a new gauge boson mass squared:
\begin{equation}
\label{deltamHu_X}
 \left(\delta m_{H_u}^2\right)_{X}= \frac{g_X^2}{64 \pi^2} m_X^2 \ln\left(\epsilon^{-2}\right) \,.
\end{equation}
Note that it depends on a parameter $\epsilon$ which also enters the effective $SU(4)$-preserving quartic coupling, cf.~eq.\eqref{eq:lambdaD}. Therefore, the tuning is typically minimized for some intermediate value of $\epsilon^2\approx\mathcal{O}(0.1)$. We show the contours of fine-tuning in the plane $\Msusy$-$g_X$ in fig.~\ref{fig:Dterm}.
We see that for the mirror TH model with $U(1)_X$ gauge symmetry the tuning may be at the level of $\mathcal{O}(10\%)$ for stop masses above 1~TeV. Another interesting feature of fig.~\ref{fig:Dterm} is that stops could be very light but in agreement with the Higgs mass measurement (even though stop mixing has been set to zero). For $\tan\beta=10$, the 125 GeV Higgs mass implies stop masses around 500~GeV.  This is in tension with the LHC direct searches for stops but heavier stops are possible, as seen from the right panel, e.g.~2~TeV stops imply $\tan\beta\approx3$.  This is in sharp contrast to the Minimal Supersymmetric Standard Model (MSSM) where, even for large $\tan\beta$, $\mathcal{O}(10)$~TeV stops are required to get the 125 GeV Higgs mass without stop mixing which results in severe fine-tuning. This follows from the fact that the tree-level Higgs mass in supersymmetric TH models is enhanced with respect to the tree-level Higgs mass in the MSSM:
\begin{equation}
\label{eq:higgsmass_tree}
 \left(m_h^2\right)_{\rm tree} \approx 2 M_Z^2 \cos^2 \left( 2\beta \right) \left(1-\frac{v^2}{f^2} \right)  \,.
\end{equation}
Since the LHC Higgs data enforces $v^2\ll f^2$ the tree-level Higgs mass is almost a factor of $\sqrt{2}$ bigger than in the MSSM.

\section{$SU(2)_X$ $D$-term Twin Higgs}

We may avoid the low Landau pole scale by generating the $D$-term potential by a non-Abelian gauge symmetry~\cite{Badziak:2017kjk}.
The matter content of the minimal model is shown in Table~\ref{tab:matter}.
In addition to the $SU(3)_c\times SU(2)_L\times U(1)_Y$ gauge symmetry and its mirror counterpart, we introduce an $SU(2)_X$ gauge symmetry which is neutral under the $\mathbb{Z}_2$ symmetry.
We embed an up-type Higgs $H_u$ into a bi-fundamental of $SU(2)_L\times SU(2)_X$, ${\cal H}$, and its mirror partner $H_u'$ into that of $SU(2)'_L\times SU(2)_X$, ${\cal H}'$. 
The $D$-term potential of $SU(2)_X$ is responsible for the $SU(4)$ invariant quartic coupling of $H_u$ and $H_u'$.
The $SU(2)_X$ symmetry is broken by the vacuum expectation value (VEV) of a pair of $SU(2)_X$ fundamental $S$ and $\bar{S}$. The relevant superpotential and soft terms is analogous to eqs.\eqref{eq:W_S}-\eqref{eq:soft_S} generalized to the case of $SU(2)_X$. The resulting $SU(4)$-invariant coupling is given by:
\begin{equation}
\label{eq:lambda_DSU2}
 \lambda=\frac{g_X^2}{8}\sin^4\beta\left(1-\epsilon^2\right) \,.
\end{equation} 
Except for $S$ and $\bar{S}$ all matter fields have their mirror partner.
The right-handed top quark is embedded into $\bar{Q}_R$ and allow for a large enough top yukawa coupling through the superpotential term ${\cal H}\bar{Q}_R Q_3$, where $Q_3$ is the third generation quark doublet. $\bar{E}$ is necessary in order to cancel the $U(1)_Y\mathchar`-SU(2)_X^2$ anomaly.
The VEV of $\phi_u$ is responsible for the masses of the up and charm quarks.
$Q_{1,2,3}$, $\bar{u}_{1,2}$, $\bar{e}_{1,2,3}$, $\bar{d}_{1,2,3}$ and $L_{1,2,3}$ are usual MSSM fields.
To cancel the gauge anomaly of 
$SU(3)_c^2 \mathchar`-U(1)_Y$ and $U(1)_Y^3$
originating from the extra up-type right handed quark in $\bar{Q}_R$ and two extra right-handed leptons in $\bar{E}$, we introduce $U$ and $E_{1,2}$.
There are three up-type Higgses in ${\cal H}$ and $\phi_u$, so we need to introduce three down-type Higgsses $\phi_{d1,2,3}$.
Their VEVs are responsible for the masses of down-type quarks and charged leptons. $SU(2)_L$ charged Higgses mix with each other by the couplings $W \sim \lambda {\cal H} \phi_d  S + \lambda {\cal H} \phi_d  \bar{S} + m \phi_u \phi_d $. More details about how the masses of the SM particles are generated can be found in \cite{Badziak:2017kjk}.

\begin{table}[htp]
\caption{The matter content of the model with an extra $SU(2)_X$ gauge symmetry.}
\begin{center}
\begin{tabular}{|c|c|c|c|c|c|c|c|}
                          &$SU(2)_X$&$SU(2)_L$&$SU(2)_L'$&$U(1)_Y$&$U(1)_Y'$&$SU(3)_c$    &$SU(3)_c'$   \\ \hline
${\cal H}$          & ${\bf 2}$    & ${\bf 2}$   &                  & $1/2$     &                &                      &                     \\
${\cal H}'$          &  ${\bf 2}$  &                  & ${\bf 2}$   &               & $1/2$       &                     &                     \\
$\bar{Q}_R$      &  ${\bf 2}$  &                  &                  & $-2/3$    &                &${\bf \bar{3}}$&                     \\
$\bar{Q}_R'$      & ${\bf 2}$   &                 &                  &               & $-2/3$     &                      &${\bf \bar{3}}$\\
$S$                    & ${\bf 2}$   &                 &                  &               &                &                      &                      \\
$\bar{S}$            & ${\bf 2}$    &                 &                  &               &                &                      &                      \\
$\bar{E}$            & ${\bf 2}$    &                 &                  & $1$        &                &                      &                      \\
$\bar{E}'$           & ${\bf 2}$    &                 &                  &               & $1$         &                      &                      \\
$U$                     &                  &                 &                  &  $2/3$     &                 & ${\bf 3}$       &                      \\
$U'$                    &                  &                 &                  &               &$2/3$         &                     & ${\bf 3}$        \\
$E_{1,2}$            &                  &                 &                  & $-1$        &                 &                      &                      \\
$E'_{1,2}$            &                  &                 &                  &               &  $-1$        &                      &                      \\
$\phi_u$              &                  & ${\bf 2}$   &                  & $1/2$     &                 &                      &                      \\
$\phi_u'$              &                  &                 & ${\bf 2}$    &               &                 &                      &                      \\
$\phi_{d1,2,3}$     &                  &${\bf 2}$   &                  & $-1/2$    &                 &                      &                      \\
$\phi_{d1,2,3}'$     &                &                 &${\bf 2}$      &               & $-1/2$   &                      &                      \\
$Q_{1,2,3}$          &                 &  ${\bf 2}$   &                 &  $1/6$     &                 &  ${\bf 3}$      &                      \\
$\bar{u}_{1,2}$     &                  &                 &                  & $-2/3$    &                 &${\bf \bar{3}}$&                      \\
$\bar{e}_{1,2,3}$ &                  &                 &                  &  $1$        &                 &                      &                      \\
$\bar{d}_{1,2,3}$ &                  &                 &                  & $1/3$      &                 &${\bf \bar{3}}$&                      \\
$L_{1,2,3}$          &                  &  ${\bf 2}$  &                  & $-1/2$    &                 &                      &                      \\
$Q_{1,2,3}'$          &                 &                 &  ${\bf 2}$   &               & $1/6$       &                      &  ${\bf 3}$       \\
$\bar{u}_{1,2}'$     &                  &                 &                  &               & $-2/3$      &                      & ${\bf \bar{3}}$ \\
$\bar{e}_{1,2,3}'$ &                  &                 &                  &               &  $1$         &                      &                      \\
$\bar{d}_{1,2,3}'$ &                  &                 &                  &               & $1/3$       &                      & ${\bf \bar{3}}$\\
$L_{1,2,3}'$          &                  &                 &  ${\bf 2}$   &               &   $-1/2$    &                      &                      \\ \hline
\end{tabular}
\end{center}
\label{tab:matter}
\end{table}%

\begin{figure}[t]
\centering
\includegraphics[clip,width=.48\textwidth]{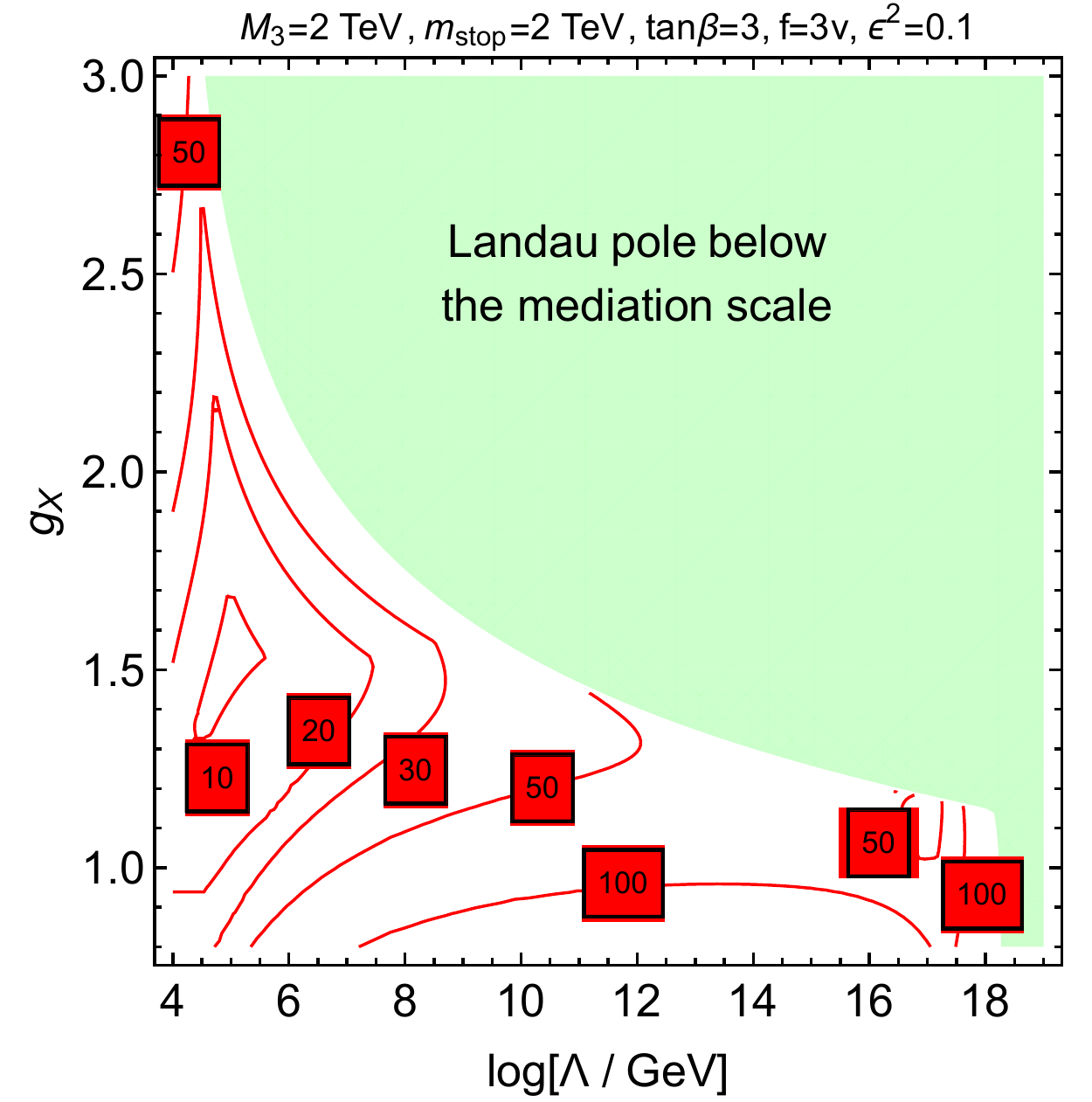}
\caption{
Fine-tuning of the $SU(2)_X$ model for $\Msusy=2$~TeV, $\tan\beta=3$ and $M_3=2$~TeV.  For the chosen values of $\Msusy$ and $\tan\beta$, the Higgs mass is in agreement with the measured value within theoretical uncertainties
in the most of parameter space. For $g_X\gtrsim1.5$ the Higgs mass is slightly too big which can be compensated by reducing $\tan\beta$ by about
10~$\%$ which would have negligible impact on fine-tuning.
}
\label{fig:gxLambda}
\end{figure}

The contours of fine-tuning in the $\Lambda\mathchar`-g_X$ plane is shown in fig.~\ref{fig:gxLambda}. The fine-tuning is not monotonously improved for larger $g_X$. This is because of the one-loop threshold correction analogous to eq.~(\ref{deltamHu_X}) (which is three times larger) as well as the two-loop RGE correction from the soft masses of $S$ and $\bar{S}$. The tuning is at the level of  $\mathcal{O}(10\%)$ for low mediation scales. Because of the slower running of the $SU(2)_X$ gauge coupling, the model remains perturbative up to the Planck scale while the tuning is at the level of few $\%$.

\section{Asymptotically free SUSY Twin Higgs}

\begin{figure}[t]
\centering
 \includegraphics[width=0.48\textwidth]{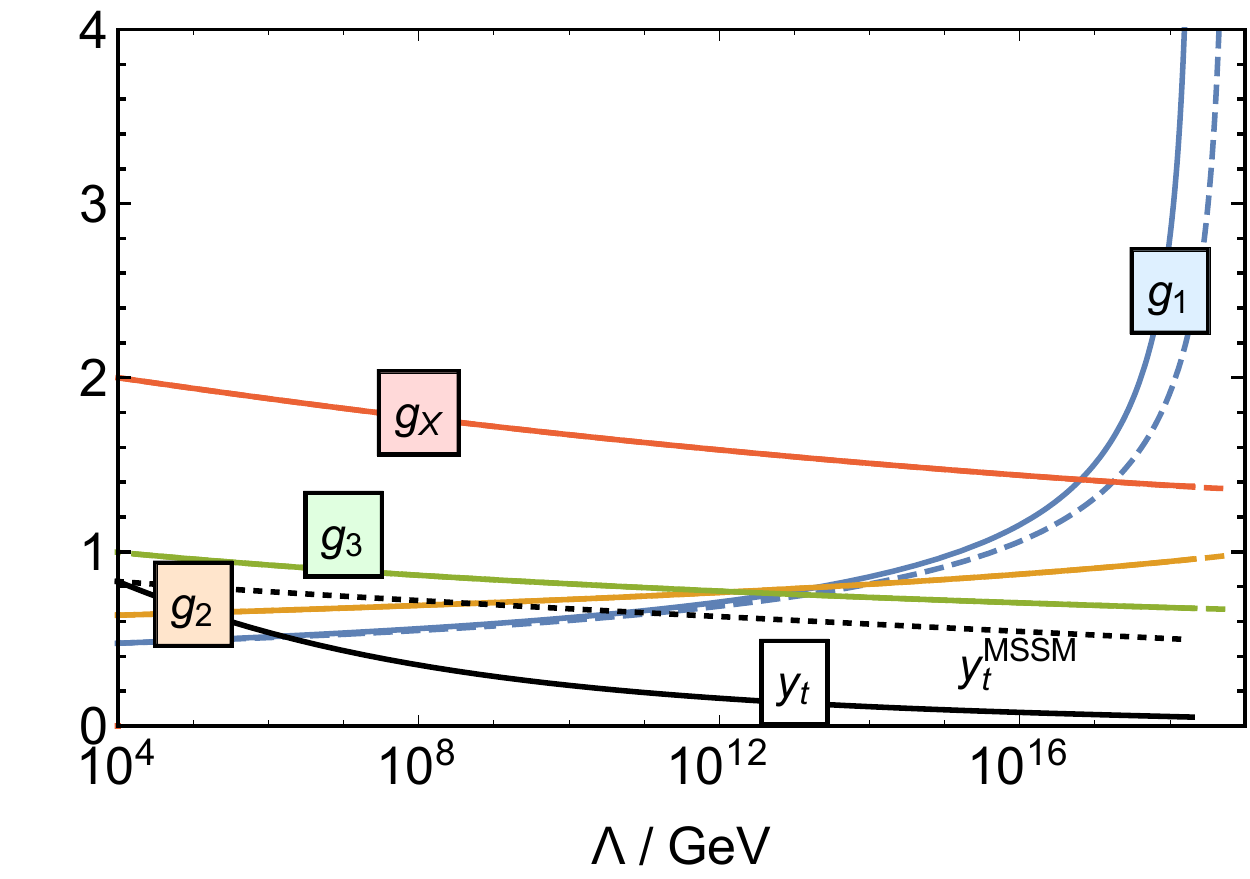}
 \caption{RG running of $g_X$ (red), $g_1$ (blue), $g_2$ (yellow), $g_3$ (green) and the top yukawa coupling $y_t$ (black) for $m_X=10$~TeV, $m_{\rm
stop}=2$~TeV, $g_X(m_X)=2$ and $\tan\beta=3$ in the model with $SU(2)_X\times SU(2)_X'$ gauge symmetry. Solid lines correspond to the case where all states beyond the Minimal Supersymmetric Standard Model
(MSSM) have masses around $m_X$. Dashed lines assume $M_{E_1}=10^7$~GeV, $M_{E_2}=10^9$~GeV, see~\cite{Badziak:2017wxn} for
details.
Dotted black line corresponds to the running
of $y_t$
in the MSSM.
}
\label{fig:running}
\end{figure}

\begin{figure}[t]
\centering
 \includegraphics[width=0.48\textwidth]{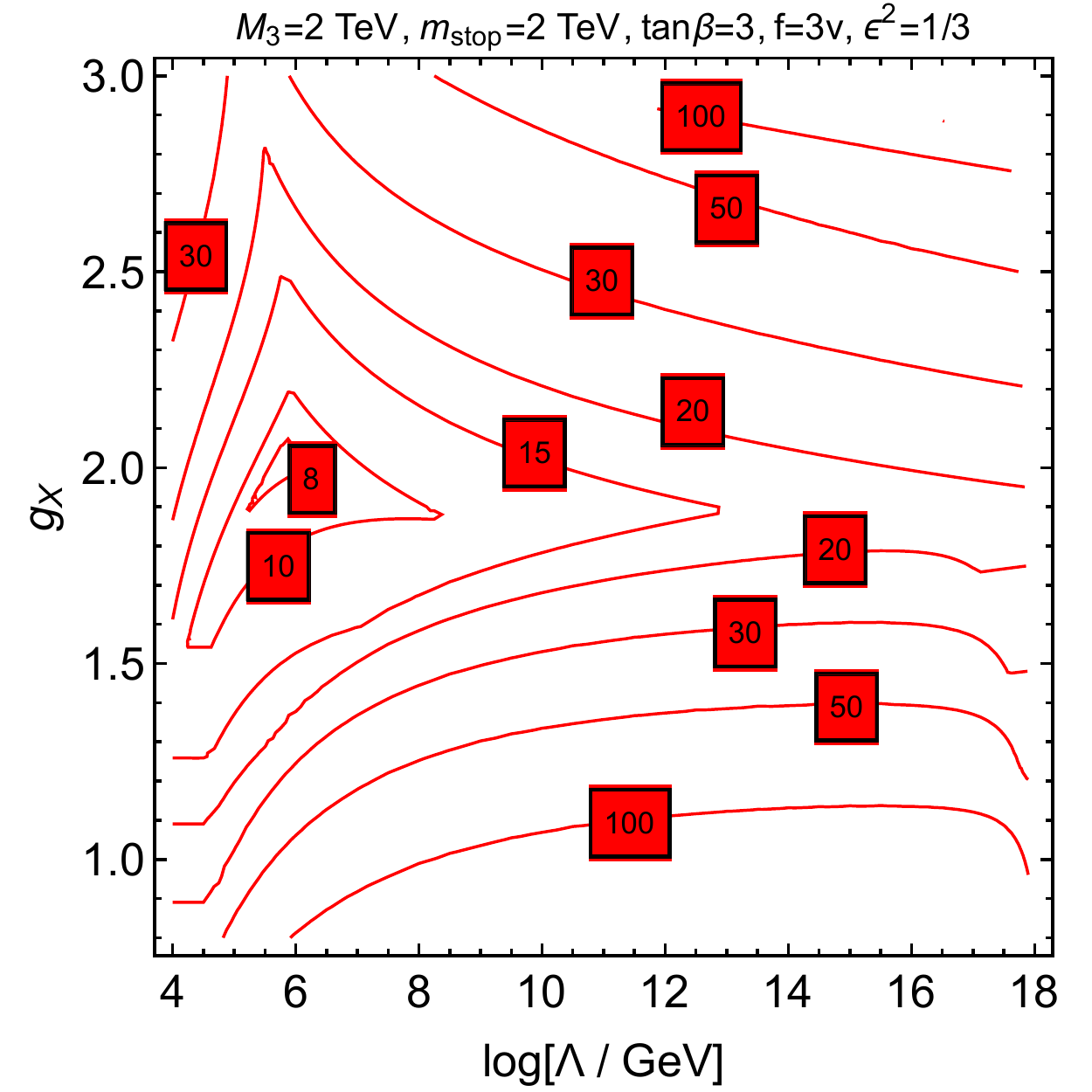}
 \caption{Fine-tuning $\Delta_v$ of the asymptotically-free model in the plane $\Lambda$-$g_X$ for  $\Msusy=2$~TeV, $\tan\beta=3$, $f=3v$, $\mu=500$~GeV,
$M_1=M_2=200$~GeV and the soft gluino mass $M_3=2$ TeV. We fix $\epsilon^2=1/3$ which corresponds to $m_S=m_X$.
}
\label{fig:ft}
\end{figure}

In the $SU(2)_X$ model, the new gauge symmetry $SU(2)_X$ is assumed to be $\mathbb{Z}_2$ neutral, and mirror particles are charged under $SU(2)_X$. We may instead consider the case where $SU(2)_X$ also has a mirror partner $SU(2)_X'$, under which mirror particles are charged. As a result the number of $SU(2)_X$ charged fields is reduced, so that the $SU(2)_X\times SU(2)_X'$ gauge interaction is asymptotically free, which is explicitly shown in fig.~\ref{fig:running}. See~\cite{Badziak:2017wxn} for the matter content and details of the mass spectrum of the model. The $SU(2)_X\times SU(2)_X'$ symmetry is broken down into the diagonal $SU(2)_D$ subgroup by the vacuum expectation value $v_\Sigma$ of a bi-fundamental field $\Sigma$ in  a supersymmetric way, e.g.~by a
superpotential $W \sim Y (\Sigma^2 - v_\Sigma^2)$ where $Y$ is a chiral multiplet, and that $v_\Sigma$ is much larger than the TeV scale, say few tens of TeV. Then below the scale
$v_\Sigma$ the theory is well-described by a SUSY theory with an $SU(2)_D$ gauge symmetry. The symmetry breaking of $SU(2)_D$ involves
SUSY breaking effect similar to model discussed in previous sections. We introduce chiral multiplets
$\Xi,~\Xi'$ and the superpotential
\begin{equation}
W = \kappa \Xi (S\bar{S}- M^2) + \kappa \Xi' (S'\bar{S}'- M^2),
\end{equation}
and soft masses,
\begin{align}
V_{\rm soft} = m_S^2 (|S|^2 + |\bar{S}|^2 +  |S'|^2 + |\bar{S}'|^2 ).
\end{align}
The diagonal subgroup is responsible for the $SU(4)$ invariant quartic coupling which is given by eq.~\eqref{eq:lambda_DSU2}.

The contours of fine-tuning in the $\Lambda\mathchar`- g_X$ plane is shown in fig.~\ref{fig:ft}. The constraint from the Landau pole of the $SU(2)_X$ interaction is absent. For 2 TeV stops and gluino the tuning is better than $5\%$ even if the mediation scale is as large as the Planck scale. The tuning is relaxed by two to three orders of magnitude as compared to the MSSM. 

It should be emphasized that the huge improvement in tuning with respect to the MSSM originates not only from the TH mechanism, guaranteed by a large $SU(4)$ invariant quartic term, and enhanced tree-level Higgs mass which allows for much lighter stops required by the 125 GeV Higgs mass. The tuning is additionally suppressed by a strong suppression of the top Yukawa coupling via RG effect of the large new gauge coupling. This suppression is demonstrated in  fig.~\ref{fig:running}. Similar effect is present also in a SUSY model without the TH mechanism and allows for tuning of $\mathcal{O}(1)~\%$ for gravity mediated SUSY breaking~\cite{Badziak:2018nnf}. 

An intriguing feature of the model is that the up quark must be embedded in the $SU(2)_X$ doublet together with the top quark to avoid the Landau pole for the hypercharge below the Planck scale. This has interesting phenomenological implications including flavor-violating top decays to the Higgs and the up quark which is potentially observable at the LHC~\cite{Badziak:2017wxn}.

\section{Summary}

We presented a new class of supersymmetric UV completions of the TH model in which the $SU(4)$ invariant quartic term is generated by a $D$-term potential of a new gauge symmetry. We showed that the Higgs mass of 125 GeV is easily obtained both for light or heavy stops. The tuning can be at the level of $\mathcal{O}(10)~\%$ for stops and gluino masses that comfortably satisfy the LHC constraints. Even if there is no sign of supersymmetry in future LHC data the tuning will remain moderate in these models. If the new gauge group is abelian a low Landau pole scale for the new interaction is required for the TH mechanism to work. However, the model can be made perturbative up to the Planck scale assuming $SU(2)_X\times SU(2)_X'$ gauge symmetry.

\section*{Acknowledgments}
MB has been partially supported by National Science Centre, Poland, under research grant no. 2017/26/D/ST2/00225. KH has been partially supported by the Director, Office of Science, Office of High Energy and Nuclear Physics, of the US Department of Energy under Contracts DE-SC0009988.

\end{document}